\documentclass[aps,prb,twocolumn,floats,epsfig,pdflatex,notitlepage]{revtex4-1}
\usepackage{color,ulem,verbatim,bbold,float}
\usepackage{amssymb,amsbsy,amsmath,mathrsfs}
\usepackage{times}
\usepackage{amssymb}
\usepackage{amsbsy}
\usepackage{amsmath}
\usepackage{float}
\usepackage{graphicx}
\usepackage{color}
\usepackage{bm}
\usepackage[plainpages=false,pdfpagelabels,colorlinks=true,linkcolor=red,urlcolor=blue,citecolor=blue,pdftitle={},pdfauthor={},pdfdisplaydoctitle={},pdfduplex=DuplexFlipLongEdge]{hyperref}
\usepackage{natbib}

\newcommand{\blu}[1] {\textcolor{black}{#1}}

\begin{document}

\title{Dynamics and correlations at a quantum phase transition beyond Kibble-Zurek}

\author{Krishanu Roychowdhury$^{1}$, Roderich Moessner$^2$, and Arnab Das$^3$}

\affiliation{$^1$Department of Physics, Stockholm University, SE-106 91 Stockholm, Sweden} 
\affiliation{$^2$Max Planck Institute for the Physics of Complex Systems, N\"{o}thnitzer Stra{\ss}e 38, 01187 Dresden, Germany}
\affiliation{$^3$Indian Association for the Cultivation of Science (School of Physical Sciences), 2A \& 2B Raja S. C. Mullick Road, Kolkata 700032, India} 

\begin{abstract}
Kibble-Zurek theory (KZ) stands out as the most robust theory of defect generation in the dynamics of phase transitions. KZ utilizes the structure of equilibrium states away from the transition point to estimate the excitations due to the transition using adiabatic and impulse approximations. Here we show, the actual non-equilibrium dynamics lead to a qualitatively different scenario from KZ, as far correlations between the defects (rather than their densities) are concerned. For a quantum Ising chain, we show, this gives rise to a Gaussian spatial decay in the domain wall (kinks) correlations, while KZ would predict an exponential fall. We propose a simple but general framework on top of KZ, based on the `quantum coarsening' dynamics of local correlators  in the supposed impulse regime. We outline how our picture extends to generic interacting situations.
\end{abstract}

\maketitle


\section{Introduction}

The Kibble-Zurek mechanism (KZM) provides arguably the simplest and most robust theory that captures the dynamics of a continuous quantum phase transition (QPT) both in the classical~\cite{Kibble_JPhys,Kibble_PhysRep,KZM_Nat,KZM_ActaPol,KZM_PhyRep} and quantum~\cite{KZM_Bogdan,KZM_ZurekZoller,Anatoli_KZM,KZM_Dziarmaga,KZM_Amit,KZM_Kris_Diptiman,Dziarmag_Rev,Kris_Rev,cincio2009dynamics} realms. Where a parameter (temperature/coupling) is ramped across the transition point at a finite rate, it predicts the universal scaling of the resulting defect density with the ramp rate (Kibble-Zurek (KZ) scaling laws). This relies on the so-called adiabatic-impulse (AI) approximation, which approximates the dynamics into two qualitatively different regimes. One is the adiabatic regime (away from the transition point), where the state evolves adiabatically with the change in the tuning parameter; the other is the impulse regime where the state of the system is considered to be effectively frozen as equilibration slows down near the critical point. The powerful simplicity of KZM continues to underpin an ever-expanding field of complex dynamics of a QPT. Universal features of the scaling 
associated with KZM have been shown to emerge for various quench protocols and extend to observables besides the local defect density, indeed even to include cases where there are no topological defects to count~\cite{KZM_chandran_sondhi,delcampouniversal,chandran2013kibble}. Remarkably, the mechanism seems to be more robust than its underlying key approximation ({\it i.e.}, the AI approximation) in predicting defect densities~\cite{KZ_Renormalized}: the dynamics in the impulse regime merely renormalizes the prefactors, leaving the scaling laws intact.

Here we pose a different question: what happens if we go beyond the ambit of the scaling hypothesis, and build on the basic elements of KZ to determine how the dynamics upon traversing the QPT affect the {\it correlations} between the defects? 
\blu{KZ, by construction, does not capture these correlations even at a qualitative level, as it does not address the dynamics of the impulse regime. Here, we supply this by considering the quantum coarsening throughout the full ramp, including the `coarsening dynamics' in the impulse regime near the QPT. Our key results are the following.
\begin{itemize}
    \item For a slow ramp, this coarsening dynamics in the impulse regime can yield unusual defect correlations,  in particular with, {\it Gaussian spatial decay}.
    \item Such correlations appear to  be absent from {\it any} eigenstate of the system. Hence, the correlations cannot be captured by any KZ-type theory relying on properties of eigenstates of any instantaneous Hamiltonian during the ramp.
    \item We propose a simple thermal picture with the kinks as the fundamental degrees of freedom, which quite accurately captures the {\it few-body defect correlations} for slow ramps. The complexity of the correlations generated by the slow but non-adiabatic ramp may be reflected in correlations involving more than a few defects, which we do not study here. 
    \item The state reached after waiting for an infinite time after the ramp concludes is described by  a novel Generalized Gibbs' Ensemble (GGE) which inherits these Gaussian correlations, while the GGE starting from the post-ramp state predicted by KZ does not.
    \item While our focus lies on extending the theory of dynamics at QPT beyond  KZ, we emphasize that our picture naturally connects with KZ; in particular, length and time-scales appearing in the novel defect correlations exhibit KZ scaling.
\end{itemize}}

The remainder of the paper is structured as follows. In Sec.~\ref{sectwo}, we outline the model of interest which is an Ising chain subjected to a transverse field that is ramped across QPTs. The correlation among the kinks/defects generated in the process is calculated in Sec.~\ref{secthree} revealing its remarkable Gaussian form. As argued in Sec.~\ref{secfour}, these correlations cannot be predicted from any eigenstate behavior, and hence, falls beyond Kibble-Zurek. In Sec.~\ref{secfive}, we propose a simple thermal picture that explains this surprising behavior of the kinks at short time scales, as well the correlations between the underlying nonlocal fermions of the theory. Sec.~\ref{secsix} demonstrates how the Gaussian correlation stabilizes at long times where a novel GGE appears, characterized by parameters inherited from the ramp wavefunctions. In Sec.~\ref{secseven}, we illuminate how our picture connects to KZ for certain length and time-scales appearing in the defect correlations exhibit KZ scaling. We conclude in Sec.~\ref{secnine} following a discussion on possible implications of our theory in real experiments in Sec.~\ref{seceight}.  


\section{Model and Protocol}\label{sectwo}

For a concrete demonstration, we consider a canonical model of QPT -- the integrable Ising chain in a transverse field ~\cite{Subir-Book, Sei-Book}. 
Our chain
\begin{equation}
 \mathcal{H}(g) = -\frac{J}{2} \sum_{j=1}^N( \sigma_{j}^{x}\sigma_{j+1}^{x}+g\sigma_{j}^{z}), 
\label{tiham}
\end{equation}
has periodic boundary $\sigma_{N+1}^x=\sigma_1^x$ and an even number of spins $N$. We set $J=1$ and ramp $g$ from $+\infty$ to $0$ over time $\tau_Q$, $g(t)=-t/\tau_Q$, crossing a critical point at $g_c=1$ between paramagnet for $|g|>1$, and ferromagnet otherwise.

We introduce dual variables $\mu^{x,z}$ ~\cite{kogut1979introduction}:
\begin{align}
 \mu_{j}^{z} &= \sigma_{j}^{x}\sigma_{j+1}^{x}~~;~~\mu_{j}^{x} = \prod_{k < j}\sigma_{k}^{z}, \label{dual} \\
\tilde{\mathcal{H}} &= - \frac{1}{2}\sum_{j} (g\mu_{j}^{x}\mu_{j+1}^{x}+\mu_{j}^{z}). \nonumber
\end{align}
Fermionization~\cite{lieb1961two,jordan1928pauli} of $\tilde{\mathcal{H}}$:
\begin{align}
 \mu_{j}^{x} = (c^\dagger_j + c_j) \prod_{l<j} (1-2c^\dagger_lc_l)~;~\mu_{j}^{z}=1-2c^\dagger_jc_j,
\end{align}
reduces this to an ensemble of two level systems in momentum space~\cite{KZM_Dziarmaga} after a Fourier transformation: $c_j = (1/\sqrt{N})\sum_j e^{i kj} c_k$ (with appropriate boundary conditions for the relevant even fermion-parity sector~\cite{damski2013exact}). 

The ground state of $\tilde{\mathcal{H}}(t)$, a tensor product $|\Psi_{\rm G}(t)\rangle\equiv\otimes_k |\Psi_k(t)\rangle$, is obtained following a time-dependent Bogoliubov-de Gennes (TDBdG) transformation: 
\begin{align}
c_k(t) = u_k(t) \gamma_k + v_{-k}^*(t) \gamma_{-k}^\dagger,
\end{align}
and demanding the state to be annihilated by the Bogoliubov fermions ($\gamma_k$) at every instant: $\gamma_k|\Psi_k(t)\rangle=0$. In Heisenberg picture, the operators satisfy~\cite{KZM_Dziarmaga} 
\begin{align}
i \frac{\rm d}{{\rm d}t} c_k=[c_k,\tilde{\mathcal{H}}]~~{\rm with}~~\frac{\rm d}{{\rm d}t} \gamma_k=0.
\end{align} 
The amplitudes $u_k(t)$ and $v_k(t)$ of $|\Psi_k(t)\rangle\equiv[u_k(t)~v_k(t)]^T$ satisfy  
\begin{equation}
 i \frac{\rm d}{{\rm d}t} |\Psi_k(t)\rangle = [\tau^z_k(1-g\cos k)+ \tau^y_k g\sin k]|\Psi_k(t)\rangle,
 \label{diffeq}
\end{equation}
where $\tau^{y,z}_k$ are Pauli matrices. Initially prepared in the ground state $|\Psi_i\rangle\equiv \otimes_k |\Psi_k(t\rightarrow-\infty)\rangle= \otimes_k [1~0]^T$, the observables at the end of the ramp are calculated via $u_k(t)$ and $v_k(t)$ obtained by solving Eq.~\ref{diffeq}. 

\begin{figure}
\begin{centering}              
\includegraphics[width=1.0\columnwidth]{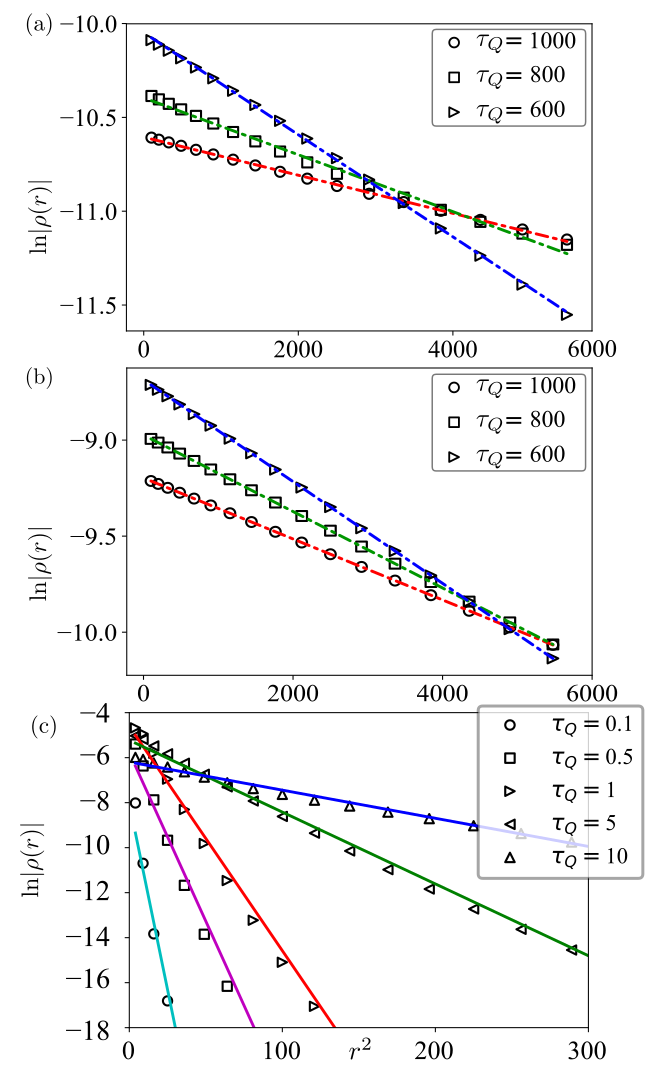}              
\end{centering}
\caption{Gaussian decay of the kink correlators. {\bf(a) Right after the Ramp:} Numerical results (for $N=10^4$) right after the ramp ending at $g=0$ (empty symbols). The dashed lines are their corresponding Gaussian fits {\bf (b) After post-Ramp Equilibration:} Showing stabilization of Gaussian behaviour after the equilibration for different values of $\tau_Q$ (ramp stopped at $g=0$). The same correlation function calculated from the GGE is shown in dashed lines (data shown for $N=10^4$. {\bf (c) Right after the Ramp -- the experimentally accessible regime:} Results for small system-sizes and rapid ramp-rate are shown. Empty symbols are numerical results and the dashed lines are their corresponding Gaussian fits (data shown for $N=50$).}    
\label{fig1}
\end{figure}


\section{Asymptotic Solution and Kink Correlator}\label{secthree}

The TDBdG can be brought to the standard form via
$|\Psi_k\rangle\rightarrow|\tilde{\Psi}_k\rangle\equiv[\tilde{u}_k~\tilde{v}_k]^T=U_k|\Psi_k\rangle$ with 
\begin{align}
U={\rm Exp}[{-i(k/2)\tau^y}]{\rm Exp}[{i(\pi/4)(\tau^z-1)}],
\end{align}
leading to the Landau-Zener-St\"{u}ckelberg type problem
\begin{align}
 i\frac{{\rm d}\tilde{u}_k(\tau)}{{\rm d}\tau} &= ~~~\frac{\tau}{\tau_Q} \tilde{u}_k(\tau) + \sin (k) ~\tilde{v}_k(\tau), \label{first:a} \\
 i\frac{{\rm d}\tilde{v}_k(\tau)}{{\rm d}\tau} &= -\frac{\tau}{\tau_Q} \tilde{v}_k(\tau) + \sin (k)~ \tilde{u}_k(\tau), \label{first:b}
\end{align}
where $\tau=t+\tau_Q\cos k$. Decoupling yields 
\begin{align}
 \bigg[ \frac{{\rm d}^2}{{\rm d}\tau^2} + \frac{i}{\tau_Q} + \frac{\tau^2}{\tau_Q^2} + \sin^2 k\bigg] \tilde{u}_k(\tau) &= 0, \label{second:a} \\
 \bigg[ \frac{{\rm d}^2}{{\rm d}\tau^2} - \frac{i}{\tau_Q} + \frac{\tau^2}{\tau_Q^2} + \sin^2 k\bigg] \tilde{v}_k(\tau) &= 0, \label{second:b}
\end{align}
and further introducing the variable $z=\tau\sqrt{2/\tau_Q}e^{-i\pi/4}$, we arrive at the Weber equations
\begin{align}
 \frac{{\rm d}^2}{{\rm d}z^2} \tilde{u}_k + \bigg[ n-\frac{1}{2}-\frac{z^2}{4} \bigg] \tilde{u}_k(z) &= 0, \label{third:a} \\
 \frac{{\rm d}^2}{{\rm d}z^2} \tilde{v}_k + \bigg[ n+\frac{1}{2}-\frac{z^2}{4} \bigg] \tilde{v}_k(z) &= 0, \label{third:b}
\end{align}
where $n=i(\tau_Q/2)\sin^2k$. It can be shown that if $P(z)$ is a solution of Eq.~\ref{third:b}, then $\big[z/2+{\rm d}/{\rm d}z\big]P(z)$ is a solution of Eq.~\ref{third:a}. The final solutions are expressed in terms of complex parabolic cylinder functions ${\cal D}_{m}(z)$~\cite{Whittaker_Watson} 
\begin{align}
 \tilde{v}_k(z) &= A {\cal D}_{-n-1}(iz) + B {\cal D}_{-n-1}(-iz), \label{fourth:a} \\
 \tilde{u}_k(z) &= \frac{e^{i\pi/4}}{\sin k\sqrt{\tau_Q/2}} \bigg[ \frac{z}{2} + \frac{{\rm d}}{{\rm d}z} \bigg]\tilde{v}_k(z), \label{fourth:b}
\end{align}
where $A, B$ are constants to be fixed by the initial conditions at $\tau\rightarrow-\infty$. The asymptotes of ${\cal D}_{m}(z)$ are given by
\begin{align}
 {\cal D}_{m}(z) &\sim e^{-z^{2}/4}z^{m}[1+\mathcal{O}(1/z^{2})], \nonumber \\ &~~~~~~~~~ \forall-\frac{3\pi}{4}< {\rm Arg}(z)< -\frac{3\pi}{4}, \\
 {\cal D}_{m}(z) &\sim \bigg(e^{-z^{2}/4}z^{m}-\frac{\sqrt{2\pi}}{\Gamma(-m)}e^{-im\pi}e^{z^{2}/4}z^{-(m+1)}\bigg) \nonumber \\ &~~~~~~~~~ \forall-\frac{5\pi}{4}< {\rm Arg}(z)< -\frac{\pi}{4}.
\end{align}
The initial conditions $\tilde{u}_{k}({\tau\rightarrow-\infty})=1$ and $\tilde{v}_{k}({\tau\rightarrow-\infty})=0$ fix the values of $A$ and $B$ as
\begin{equation}
A = 0~~;~~B = \sin k\sqrt{\frac{\tau_Q}{2}} e^{-\pi\tau_Q\sin^{2}k/8}.
\end{equation}
We use the above asymptotes for $\tau\rightarrow \infty$, which for the ramps ending at $g=0$ is ensured by $\tau_Q\gg1$ (i.e. slow ramps), to calculate $\tilde{u}_k$ and $\tilde{v}_k$. Thus, at the end of the ramp ($t=0$),
\begin{align}
\tilde{u}_k=r_k e^{i\omega_k}~~{\rm and} ~~\tilde{v}_k={\rm sgn}(\Delta)\sqrt{1-r^2_k} e^{i\phi_k},
\end{align}
with
\begin{align}
r_k &= e^{-\pi\tau_Q\Delta^2/2},~\omega_k = \frac{3\pi}{4}-\frac{\tau^2}{2\tau_Q}-\frac{\tau_Q\Delta^2}{2}{\rm ln}\bigg(\tau\sqrt{2/\tau_Q}\bigg) \nonumber \\ 
\phi_k &= \frac{\tau^2}{2\tau_Q} + \frac{\tau_Q\Delta^2}{2}{\rm ln}\bigg(\tau\sqrt{2/\tau_Q}\bigg)-{\rm arg}(\Gamma[1+i\tau_Q\Delta^2/2]),
\end{align}
where $\Delta=\sin(k)$ and $u_k,v_k$ follow from $|\Psi_k\rangle=U^\dagger|\tilde{\Psi}_k\rangle$ as
\begin{align}
 u_k &= \cos (k/2) \tilde{u}_k + \sin (k/2) \tilde{v}_k \nonumber \\ 
 v_k &= -i \sin (k/2)\tilde{u}_k + i \cos (k/2) \tilde{v}_k.
 \label{ukvk}
\end{align}

The correlators of primary concern are those between kinks in the ferromagnetic phase (of the original model, Eq.~\ref{tiham}) and generated in the course of the ramp in the vicinity of the critical point. The kinks are topological defects deep in the ferromagnetic phase where the ramp ends (at $t=0$) with their number scaling with the ramp time $\tau_Q$ as $n_d\sim\tau_Q^{-1/2}$~\cite{KZM_ZurekZoller,KZM_Dziarmaga}. 

Our first central result is this: the kink correlators feature an unusual {\it Gaussian decay in space} instead of the more familiar exponential or power-law behavior. The correlator in the longitudinal direction is of the form $\hat{\rho}(r)=(1-\sigma^x_j\sigma^x_{j+1})(1-\sigma^x_{j+r}\sigma^x_{j+r+1})$ and the corresponding connected correlator is
\begin{align}
\rho(r)=\langle\hat{\rho}(r)\rangle-\langle1-\sigma^x_j\sigma^x_{j+1}\rangle^2 = \langle\mu_j^z\mu_{j+r}^z\rangle-\langle\mu_j^z\rangle^2, 
\end{align}
where the duality relations in Eq.~\ref{dual} are invoked along with translation invariance. \blu{In terms of hardcore bosons $b_{i}(b_{i}^{\dagger}) = [\mu_{i}^{x} +(-)~i\mu_{i}^{y}]/2,$ $\mu_{i}^{z} = 1-2b_{j}^{\dagger}b_{j}= 1-2n_{j}$, which express $\rho(r)$ in terms of bosonic density-density correlation:
$\rho(r) = 4[\langle n_{j}n_{j+r} \rangle - \langle n\rangle^{2}]$ where a uniform boson density is assumed owing to translational invariance {\it i.e.}, $\langle n \rangle = \langle n_{j}\rangle$ for any $j$}. The expectation is with respect to the (time-independent) ground state of the Bogoliubov fermions \blu{(these kinks serve as the defects for $|g| <1$)}.
Using fermionization, the correlation reads
\begin{align}
 \rho(r) &= \frac{4}{N^2}\bigg[ \sum_{k>0} |v_k|^2 e^{-ikr}\sum_{l>0} |u_l|^2 e^{-ilr} \nonumber \\
 &~~~~~~~~~~~~~~~~~~~~~~~~~~~~~~~~~~~~~~~~~~ -\Big|\sum_{k>0} u_k v_k^* e^{-ikr}\Big|^2 \bigg].
\label{corrukvk}
\end{align}
Using the expressions of $u_k$ and $v_k$ noted before, 
\begin{align}
 |u_k|^2 &= \sin^2(k/2) + r_k^2 \cos k + r_ks_k \sin k \cos(\phi_k-\omega_k) \nonumber \\
 |v_k|^2 &= \cos^2(k/2) - r_k^2 \cos k - r_ks_k \sin k \cos(\phi_k-\omega_k) \nonumber \\ 
 u_kv_k^\ast &= i\frac{\sin k}{2}(r_k^2-s_k^2) - ir_ks_k \cos k \cos(\phi_k-\omega_k) \nonumber \\
 &~~~~~~~~~~~~~~~~~~~~~~~~~~~~~~~~~~~~~~~~~~~ - r_ks_k \sin(\phi_k-\omega_k),  
\end{align}
with $s_k={\rm sgn}(\Delta)\sqrt{1-r^2_k}$, from which a simple analytic form of the correlation function can be achieved ignoring the terms with rapidly oscillating phases $e^{i\omega_k}$ and $e^{i\phi_k}$ for $\tau_Q\gg1$. A straightforward calculation gives in the limit $N\to\infty$
\begin{equation}
 \rho(r) \approx -\frac{1}{4\pi^2 \tau_{Q}}e^{-r^2/2\pi\tau_{Q}},
\label{corr1}
\end{equation}
for $r\gg1$ with the sum over $k$ being replaced by integrals as $(2/N)\sum_{k>0}()\rightarrow (1/\pi)\int_0^\pi~{\rm d}k~()$. This is verified numerically in Fig.~\ref{fig1} (a). 

Such Gaussian form of the kink correlator is observed also for a full ramp from $g=+\infty$ to $g=-\infty$; as well as in the {\it transverse} correlator $(1-\sigma^z_j\sigma^z_{j+1})(1-\sigma^z_{j+r}\sigma^z_{j+r+1})$ both for half and full ramps as shown in Appendix~\ref{app:A}.


\section{Gaussian Correlations: Departure from Eigenstate Correlations and KZ}\label{secfour}

The Gaussian correlation emerges essentially out of the complex dynamics of slow ramps, and cannot be derived simply from the knowledge of the eigenstates of the static Hamiltonian, and hence from the KZ picture. Recall that KZ assumes that the system remains in its ground state until at $t = +\hat{t}$, where it falls out of equilibrium and effectively stops evolving. This way, the properties of the equilibrium state ({\it i.e.}, the ground state) in one phase is used for obtaining excitations in the other phase. It is, hence, imperative to ask whether the Gaussian decay after the ramp is likewise inherited from the ground state of the Hamiltonian ${\cal H}(g)$ (Eq.~\ref{tiham}) for some $g$ in the initial phase.

The answer is no: the ground state correlations decay exponentially for any $|g|\neq g_c$. In detail, in the ground state of ${\cal H}(g)$, 
\begin{align}
 \rho(r) = -{G}(r){G}(-r) ~~;~~ G(r) = (1/\pi)\int_0^\pi {\rm d}k~{\cal G}
\end{align}
with
\begin{align}
 {\cal G} =  [\cos(kr)-g\cos (kr-k)]/\sqrt{1-2g\cos k+g^2}.
 \label{calG}
\end{align}
The integral involves regularized hypergeometric functions whose asymptotes at large values of $r$ yield the falling exponential $G(r)\sim e^{-r/\xi_0}$ with the correlation length $\xi_0$ diverging at $g=g_c$ which is worked out in detail in Appendix~\ref{app:B}. 

In fact our investigation suggests that the non-Gaussian behavior is generic also to excited eigenstates of all 
sorts of energy densities. One family of such (excited) eigenstates can be prepared by altering the occupation of 
all $k$-modes up to a certain value $k^*\in[0,\pi]$~\cite{nandy2016eigenstate} ($k^*=0$ implying the ground state and $k^*=\pi$ implying the highest excited state) corresponding to an energy density 
\begin{align}
 \bar{E}=(2/\pi)\int_{0}^{k^*}{\rm d}k~\sqrt{1-2g\cos k+g^2}.
 \label{Ebox}
\end{align}
The quantity $G(r)$, in this case, takes the form 
\begin{align}
 G(r) = (1/\pi) \int_0^\pi {\rm d}k~{\rm sgn}(k-k^\ast){\cal G}.
 \label{boxG}
\end{align}
Any excited state prepared by occupying a randomly selected set of $k$-modes in $[0,\pi]$ with a given probability density function would also contain the same algebraic factors as in ${\cal G}$, and so, the resultant correlation function cannot display any Gaussian--unless a Gaussian factor is explicitly present already in ${\cal G}$ which, of course, is the case for the ramp owing to the functional forms of $u_k$ and $v_k$ noted previously. This eliminates possible AI type  explanations of the Gaussian spatial decay, as the evolution of the wave-function away from the ground states needs to be taken into account. 
\blu{In particular, in the simplest KZ picture, the kink-kink correlations at the critical point would be those of the ground state at the first adiabatic-impulse boundary, hence, will exhibit an {\it exponential decay} rather than the Gaussian. The same conclusion can be shown to hold for the ramp continued to infinity. In that case, instead of the observed Gaussian, AI yields $\rho(r)\sim e^{-r/\xi_0}$ as illustrated below}.

In AI approximation, the evolution of the wavefunction starting at $t=t_i\ll -\hat{t}$ till $t=t_f\gg \hat{t}$ is split into three stages: (i) adiabatic for $t\in [t_i, -\hat{t}]$, (ii) impulse for $t\in [-\hat{t}, \hat{t}]$, and (iii) adiabatic again for $t\in [\hat{t}, t_f]$ ($\hat{t}$ is a phenomenological parameter that shows a scaling with $\tau_Q$ predicted by KZM). The full time evolution operator is 
\begin{align}
 U(t) = \otimes_{k>0} U_k(t)~;~ U_k(t_f,t_i) = {\cal T}{\rm exp}\bigg(-i\int_{t_i}^{t_f} \tilde{\cal H}_k(t)~{\rm d}t  \bigg),
\end{align}
where $\tilde{\cal H}_k(t)=\tau^z_k(1-g\cos k)+ \tau^y_k g\sin k$, and ${\cal T}$ denoting time-ordering. During the adiabatic evolution, we can approximate the time evolution operator at each $k$ as~\cite{kato1950adiabatic, mostafazadeh1997quantum, tomka2018accuracy}
\begin{align}
 U^{(\rm ad)}_k(t_a,t_b) = \sum_{n=0,1} e^{i\alpha_k^{(n)}(t_a,t_b)}|n_k(t_b)\rangle\langle n_k(t_a)|,
\end{align}
with
\begin{align}
 \alpha_k^{(n)}(t_a,t_b)=-\int^{t_a}_{t_b} E_k^{(n)}(t)~{\rm d}t + i\int^{t_a}_{t_b} \langle n_k(t)|\dot{n}_k(t)\rangle ~{\rm d}t,
\end{align}
where $|n_k(t)\rangle$ denotes an instantaneous eigenstate of $\tilde{\cal H}_k(t)$ with eigenvalue $E_k^{(n)}(t)$ ($n=0$ implying the ground state).

Following AI approximation, KZ predicts the resulting wavefunction at $t=t_f$ to be of the form $|\psi^{(\rm AI)}(t_f)\rangle=\prod_k |\psi^{(\rm AI)}_k(t_f)\rangle$ where
\begin{align}
 |\psi^{(\rm AI)}_k(t_f)\rangle = U^{(\rm ad)}_k(t_f,\hat{t})~\hat{1}~U^{(\rm ad)}_k(-\hat{t},t_i)|\psi_k(t_i)\rangle,
\end{align}
the initial state at $t=t_i$ being $|\psi(t_i)\rangle=\prod_k |\psi_k(t_i)\rangle$. Assuming this initial state to be the ground state of $\tilde{\cal H}_k(t)$ at $t=t_i$ for each $k$: $|\psi(t_i)\rangle=|0
_k(t_i)\rangle$, 
\begin{align}
 |\psi^{(\rm AI)}_k(t_f)\rangle &= \sum_{n} e^{i\alpha_k^{(n)}(t_f,\hat{t})} e^{i\alpha_k^{(0)}(-\hat{t},t_i)} \nonumber \\
 &~~~~~~~~~~~~~~~~~~~~~~ \times\langle n_k(\hat{t})|0_k(-\hat{t})\rangle |n_k(t_f)\rangle. 
\end{align}
The kink correlation in AI approximation is given by $\rho^{(\rm AI )}(r) = -{G}^{(\rm AI )}(r){G}^{(\rm AI )}(-r)$ where
\begin{align}
 G^{(\rm AI)}(r) = \frac{2}{N}\sum_{k>0} \cos[\theta_k(\hat{t})-\theta_k(-\hat{t})]{\cal G},
\end{align}
($\mathcal{G}$ defined previously in Eq.~\ref{calG}), where we have used 
\begin{align}
 &\langle n_k(\hat{t})|0_k(-\hat{t})\rangle = \nonumber \\ &\cos\bigg(\frac{\theta_k(\hat{t})-\theta_k(-\hat{t})}{2}\bigg)\delta_{n,0} + \sin\bigg(\frac{\theta_k(\hat{t})-\theta_k(-\hat{t})}{2}\bigg)\delta_{n,1}.
 \label{alg}
\end{align}
Defining $g(t=\hat{t})\equiv \hat{g}$ and noting $g(t=-\hat{t})=-\hat{g}$,  
\begin{align}
 \cos[\theta_k(\pm\hat{t})] &= \frac{1\mp \hat{g}\cos k}{\sqrt{1+\hat{g}^2 \mp 2 \hat{g}\cos k}}, \nonumber \\ 
 \sin[\theta_k(\pm\hat{t})] &= \frac{\pm \hat{g}\sin k}{\sqrt{1+\hat{g}^2 \mp 2 \hat{g}\cos k}},
\end{align}
which yield
\begin{align}
 G^{(\rm AI)}(r) = \frac{2}{N}\sum_{k>0} \frac{1-\hat{g}^2}{\sqrt{(1+\hat{g}^2)^2 - 4 \hat{g}^2\cos^2 k}} {\cal G}.
\end{align}
Note $0\le 4\hat{g}^2/(1+\hat{g}^2)^2\le 1$, and so, we can approximate
\begin{align}
 G^{(\rm AI)}(r) \approx \frac{1- \hat{g}^2}{1+\hat{g}^2} \Bigg(\frac{2}{N}\sum_{k>0} {\cal G}\Bigg) \sim e^{-r/\xi_0},
 \label{aiaexpon}
\end{align}
leading to an exponentially decay of the kink correlation $\rho^{(\rm AI)}(r)$ resembling its eigenstate behavior. 


\section{Theory of defect correlations}\label{secfive}

\blu{We next show how the coarsening dynamics in the impulse regime changes the  qualitative picture for the defect correlations. Additionally, some of the resulting few-body correlations turn out to be captured well by an effective thermal description}.

Gaussian decay of correlations is not common in isolated many-body quantum systems. However, it is observed in dilute gas of local bosons in low dimensions, including trapped ideal (see, e.g.~\cite{Hadzibabic_Rev, naraschewski1999spatial, wright2012two}) or repulsive Bose gases~\cite{sykes2008spatial}. For a non-degenerate Bose gas at high temperature ($T\gg T_{\rm QD}$, $T_{\rm QD} =$  quantum degeneracy temperature), the density correlations are Gaussian in the classical limit, crossing over to exponential in the quantum limit ($T\ll T_{\rm QD}$). \blu{This does indeed resemble the behaviour of our {bosonic} kinks, which likewise in the slow ramp limit ($\tau_Q\gg 1$), exhibit Gaussian correlations}. 

For spin correlations, the relevant degrees of freedom (fermions) are {non-local in space} (see, e.g.,~\cite{Sei-Book}). Here, our thermal picture automatically accounts for the long-known oscillatory behavior of the spin correlations at the end of a ramp across the QPT in an Ising chain~\cite{Cherng_Levitov}. Unlike the kink correlators, the spin correlators $\langle \sigma^x_i\sigma^x_{i+r}\rangle$, involve nonlocal fermions, and their two-point correlators $\langle c_i^\dagger c_{i+r}\rangle$ and related observables are influenced by the formation of algebraic kink-antikink features on a length-scale set by the Fermi wavelength, with an exponential envelope due to the finite temperature. These oscillations are prominent at large ramp times ({\it i.e.}, slow ramps), but vanish abruptly below a threshold value, $\tau_Q^*$~\cite{Cherng_Levitov}. In detail, we find a thermal occupancy with $T=1/(\pi\tau_Q)$ and dispersion $\varepsilon_k=\tau_Q^*/\tau_Q-\sin^2k\le0$. The pair of Fermi points present for $\tau_Q>\tau_Q^*$ merge at zero wave vector when $\tau_Q=\tau_Q^*$, yielding a gap for $\tau_Q<\tau_Q^*$ and terminating the oscillatory behavior. Further details are provided in Appendix~\ref{app:C}. 

\blu{Applying a thermal picture -- rather than a GGE -- to a post-ramp state in an {\it integrable} system seems like a rather drastic approximation. Its accuracy can nonetheless be demonstrated numerically, (a) for slow ramps, so the kinks are well separated and hence only weakly interacting, and  (b) for {\it few-body} operators. Indeed, considering these, it is often rather hard to distinguish an integrable GGE from a standard (thermal) Gibbs ensemble~\cite{Rigol_GGE_vs_GE_1, Rigol_GGE_vs_GE_2, Rigol_GGE_vs_GE_3}, a distinction which becomes more apparent for more complex correlations~\cite{Rigol_GGE_vs_GE_4}}. As it turns out, at short times, the behaviour of the kinks indeed appear thermal which can be demonstrated by computing the kink density and kink correlation in a thermal ensemble characterized by an effective temperature that corresponds to the energy density produced in the final state at the end of the ramp.

The average energy of a system of free kinks maintained at inverse temperature $\beta$ is given by the equipartition theorem $E_{\rm av} =(2\beta)^{-1}$ which we equate with the energy $\delta E \equiv \langle \psi_{\rm fin}|{\cal H}(g)|\psi_{\rm fin}\rangle - E_0$ at the end of the ramp $g=0$ ($E_0$ is the ground state energy of ${\cal H}(g)$ at $g=0$ and the final state thereof is denoted by $|\psi_{\rm fin}\rangle\equiv
\begin{pmatrix}
u_k & v_k 
\end{pmatrix}^T$, calculated numerically). Taking a unit mass of the kinks ($M=1$), this reveals a scaling of $\beta$ as $\beta\sim \sqrt{\tau_Q}$. However, the actual parameter that enters the calculations for kink density and higher-order correlators is $\beta/M$ -- this parameter scales with $\tau_Q$, implying a possible $\tau_Q$-dependent scaling of the mass term as well. 

The kink density in terms of $\beta$ is given by 
\begin{align}
 n_{\rm kink}(\beta) = \frac{1}{N} \sum_{k} e^{-\beta\epsilon_k} \approx \sqrt{\frac{M}{2\pi\beta}}
\end{align}
for $\epsilon_k=k^2/2M$. Exact calculations~\cite{KZM_Dziarmaga} suggest that the transverse magnetization $m_z=\langle \sigma^z\rangle-\langle \sigma^z\rangle_0$ ($\langle \sigma^z\rangle_0$ denoting the saturation value) calculated within AI scales as $m_z \sim 1/\sqrt{\tau_Q}$ implying $\beta/M\sim \tau_Q$ as this quantity can be identified with the density of kinks ($n_{\rm kink}$) calculated above. 
Including the scaling for $\beta\sim\sqrt{\tau_Q}$, we obtain $M\sim 1/\sqrt{\tau_Q}$. In other words, the effective temperature for the kinks should be identified with $M/\beta$ that scales with $\tau_Q^{-1}$, similar to what holds also for the underlying (nonlocal) fermions in our theory. 

The density-density correlation among the kinks with the abovementioned scaling of their mass is likewise given by
\begin{align}
 \rho_{\rm d-d}(r) &= \langle n_i n_{i+r} \rangle - \langle n_i\rangle \langle n_{i+r} \rangle \nonumber \\
 &= -G(r)G(-r) \sim \frac{1}{\tau_Q}e^{-r^2/\tau_Q},
\end{align}
as
\begin{align}
 G(r) \approx \frac{1}{2\pi}\int_{-\infty}^\infty{\rm d}k~\cos(kr) e^{-(\beta/2M)k^2} = \sqrt{\frac{M}{2\pi\beta}} e^{-Mr^2/2\beta},
 \label{ddcor}
\end{align}
and shows a Gaussian behavior which matches with the previously calculated kink correlation $\rho(r)$ following the ramp. The correlation length in Eq.~\ref{ddcor} is essentially the thermal de Broglie wavelength $\lambda_{\rm th}\sim \sqrt{\beta/M}$ for the kinks whose scaling with $\tau_Q$ can be identified with that of the Gaussian decay due to the ramp, {\it viz}., $\lambda_{\rm th}\sim \sqrt{\tau_Q}$.

Our theory should be robust even in generic systems, {\it i.e.}, in presence of (weak) interactions: neither of its two necessary elements -- small density of defects at slow ramps, and an effective thermal environment -- relies on the absence of interaction/integrability. In fact, a thermal local environment is to be expected from the eigenstate thermalization hypothesis (ETH) in an interacting system driven out of equilibrium~\cite{Jaynes, Srednicki, rigol2007relaxation, cassidy2011generalized, rigol2008thermalization, bordia2017periodically}. Moreover, a small residual interaction between the bosonic excitations does not affect the Gaussian behavior at large separations, but only modifies it at small distances with corrections of the form $\delta\rho(r) = -(2\gamma/nr)(T/T_{\rm QD})^{-1}~{\rm Exp}[-(n^2r^2/2)(T/T_{\rm QD})]$ where $\gamma$ and $n$ correspond to the interaction strength and particle density respectively~\cite{sykes2008spatial}. This also exhibits Gaussian behavior at large $r$.


\section{Equilibration of Correlations After stopping the Ramp and resulting GGE}\label{secsix}

The Gaussian behavior is observed to persist in the long-time limit after the ramp is stopped and the system is allowed to evolve with a time-independent final Hamiltonian frozen at $g=0$ 
[Fig.~\ref{fig1}(b)]. At late times, this leads to the expressions (see Appendix~\ref{app:D} for details)
\begin{align}
 |u_k|^2 &\approx \frac{1}{2}\big(1+\cos^2\theta_k\big) \big|{u}_k^{(\rm I)}\big|^2 + \frac{1}{2}\big(1-\cos^2\theta_k\big) \big|{v}_k^{(\rm I)}\big|^2 \nonumber \\
 |v_k|^2 &= 1-|u_k|^2 \nonumber \\
 u_kv_k^\ast &= \frac{i}{2}\sin\theta_k\cos\theta_k\bigg(\big|{v}_k^{(\rm I)}\big|^2-\big|{u}_k^{(\rm I)}\big|^2\bigg),
 \label{ukvkpostramp}
\end{align}
where ${u}_k^{(\rm I)}$ and ${v}_k^{(\rm I)}$ are the parameters for the initial state obtained immediately after the ramp stopped at $g=0$ and given in Eq.~\ref{ukvk}. With these, the final behavior of the kink correlation (Eq.~\ref{corrukvk}) in the long-time limit is shown in Fig.~\ref{fig1} (b). 

This long-time dynamics amounts to a GGE with the density matrix
$\varrho_{\rm GGE} = Z^{-1} e^{-\sum_k \lambda_k J_k} 
\left(Z={\rm Tr}[e^{-\sum_k \lambda_k J_k}]\right)$ and $\lambda_k = {\rm ln}[(1-\langle \psi_0|J_k|\psi_0\rangle)/\langle \psi_0|J_k|\psi_0\rangle]$. 
Here the information of the ``initial state" (reached at the end of the ramp) is encoded in the Lagrange multipliers $\lambda_{k}$ corresponding to the conserved quantities $J_k=\gamma^\dagger_k\gamma_k$~\cite{vidmar2016generalized}. The details on computing $\varrho_{\rm GGE}$ and the kink correlation function ${\rho}_{\rm GGE}(r)$ using $\varrho_{\rm GGE}$ are presented in Appendix~\ref{app:E} and \ref{app:F}. Here we state the final results:
\begin{align}
{\rho}_{\rm GGE}(r) = &-\frac{4}{N^2}\sum_{k>0} (2\alpha_k-1)^2\mathcal{G}(r)\mathcal{G}(-r) \nonumber \\
&+ \frac{4}{N^2} \sum_{k>0} 2\alpha_k(1-\alpha_k)\sin^2 (kr),
\label{rhoGGE}
\end{align}
where $\mathcal{G}(r)$ is given in Eq.~\ref{calG} and $\alpha_k$ is the overlap between the wavefunction at $k$ obtained immediately after stopping the ramp at $g=0$ (denoted as the initial state $[u_k^{(I)}, v_k^{(I)}]^T$) and that obtained by diagonalizing $\tilde{\cal H}(g)$ at $k$ for $g=0$ (denoted as the final state $[u_k^{(F)}, v_k^{(F)}]^T$),
\begin{align}
 \alpha_k \approx \big|u_k^{(\rm I)}\big|^2\big|u_k^{(\rm F)}\big|^2+\big|v_k^{(\rm I)}\big|^2\big|u_k^{(\rm F)}\big|^2,
\end{align}
where we have neglected the terms that contain the rapidly oscillating phases $e^{i\omega_k}$ and $e^{i\phi_k}$ for $\tau_Q\gg1$ as also done previously. \blu{Evidently, the Gaussian decay is inherited from the state produced by the ramp and imprinted in the Lagrange multipliers. We note that Gaussian correlations are absent in a GGE that arises due to an instantaneous quench from the {\it ground state or any other eigenstate} of ${\cal H}(g)$ for any value of $g$: the complexity of the coarsening dynamics is crucial for their genesis}.

In non-integrable systems, our picture will qualitatively concur with  KZ  at  long times after the ramp ends: here, a genuine thermal description should apply, for whose temperature KZ scaling could provide an estimate. By contrast, KZ would not naturally predict Gaussian correlations at `short' time, {\it i.e.}, in the state right at the end of the ramp. There we, however, also expect to observe them as {\it removing} the constraints imposed by integrability should not hinder their build-up. 


\section{Consistency with KZ}\label{secseven}

\begin{figure}
\begin{centering}
\includegraphics[width=1.0\columnwidth]{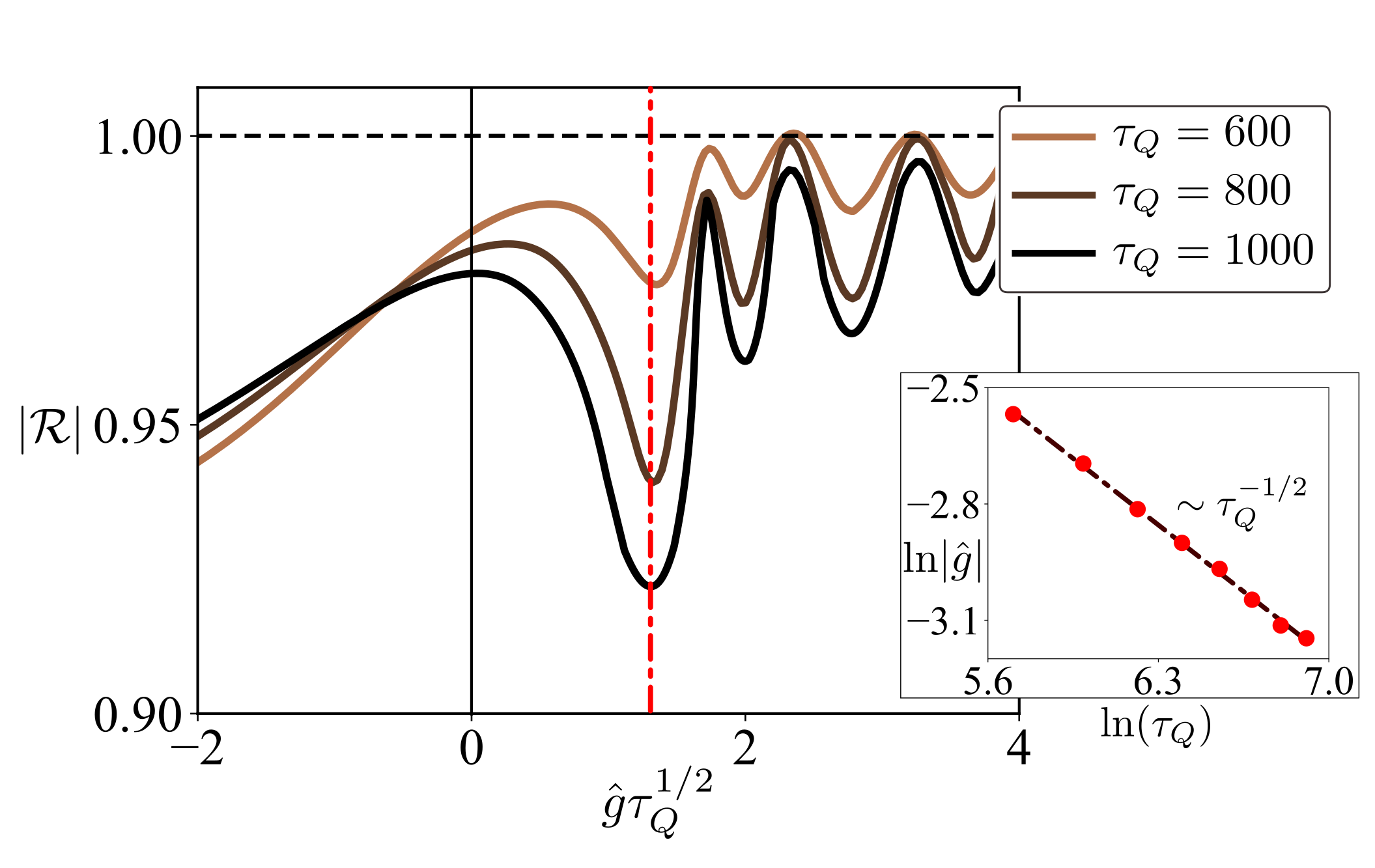}              
\end{centering}
\caption{{\bf KZ scaling of the set-in time for the Gaussian decay:} (a) Pearson correlation coefficient $|\mathcal{R}|$ versus KZ-scaled $\hat{g}$ (defined in the text) revealing the extrema coincide (the primary one guided by the red dashed line) when $\hat{g}\rightarrow \hat{g}\tau_Q^{1/2}$. {\bf Inset:} Set-in time $\hat{g}$ of the Gaussian behavior against $\tau_Q$ confirming KZ scaling $\hat{g}\sim\tau_Q^{-1/2}$ (data shown for $N=10^4$).}
\label{fig2}
\end{figure}

While the form of the correlations are beyond the KZ framework, it nonetheless accounts for the scaling of most of the relevant length and time scales, reflecting the critical slowing down as a central ingredient~\cite{DZS_Winding}.

We begin with Eq.~\ref{corr1}, where it is the scale of the Gaussian decay: $\xi\sim\tau_Q^{1/2}$ which reflects KZ scaling~\cite{KZM_Dziarmaga}. Also, Gaussian behavior sets in at $g(t) = g^{\ast}$ which in turn obeys KZ scaling: $\hat{g} = (g^{\ast}-g_{c})/g_{c}\sim \tau_Q^{-1/2}$. As a heuristic for determining the set-in point $\hat{g}$, we consider estimating the (absolute value of the) Pearson correlation coefficient ($|{\cal R}|\in[0,1]$) for the plot of ln$|\rho(r)|$ against $r^2$ (for details, see Appendix~\ref{app:G}) as a function of $\hat{g}$ for different values of $\tau_Q$. {{For a perfectly linearly correlated data-set $(X=\{x_i\},Y=\{y_i\})$, $|\mathcal{R}|$ is $1$. Accordingly, if the data-set is perfectly Gaussian correlated, $|\mathcal{R}|$ is $1$ for $X=\{x_i^2\}$ and $Y=\{{\rm ln}|y_i|\}$. Following this, we compute $|\mathcal{R}|$ for the data-set $\mathcal{D}:=\{(r_i^2, {\rm ln}|\rho_i|)\}$}} and mark the location of the first minimum (from the critical point $g_c=1$) which reveals KZ scaling of $\hat{g}$: $\hat{g}\sim \tau_Q^{-1/2}$ (Fig.~\ref{fig2} inset). For illustration, in Fig.~\ref{fig2} main, we plot the behavior of $\mathcal{R}$ against the KZ-scaled $\hat{g}$ such that the extrema for different values of $\tau_Q$ coincide.


\section{Near-time Experimental Realm}\label{seceight}

The novel correlations should be accessible experimentally using programmable Rydberg atomic quantum simulators~\cite{KZ_Exp_Keesling_Lukin} or Ising model simulators using superconducting qubits~\cite{KZ_Exp_China}. KZM has already been simulated using both setups. In Fig.~\ref{fig1} (c), we show the characteristic behavior for parameter ranges well below the analytical requirement $\tau_{Q}\gg 1$. For comparison, the regime accessed by experiments is $\tau_{Q} \approx 14$ for a many-body Rydberg atom system~\cite{KZ_Exp_Keesling_Lukin} and $\tau_{Q} \approx 620$ for a superconducting qubit simulator~\cite{KZ_Exp_China}. Note that, for $\tau_{Q}=5$,  the value of the short-distance correlations, due to the increased initial defect density, is considerably enhanced [Fig.~\ref{fig1} (c)], and thence the correlation decay quite accessible at short distances.


\section{Conclusion}\label{secnine} 

\blu{Our analysis of dynamics and correlations beyond the AI scenario of KZ endows the `inert' impulse regime with a quantum coarsening dynamics of its own. Following  the provision of a nontrivial initial state generated by the ramp, we are led to a novel GGE, apparently naturally accessible only via such a route.
This represents a departure from the KZ philosophy, which is to estimate non-equilibrium properties in the ordered phase entirely from the structure of a suitably chosen {\it ground state} in the disordered phase, as the system is ramped from the latter to the former. Our picture is also attractive as the coexisting qualitatively distinct (and unusual) correlations can be rationalized via thermal descriptions of the underlying respective fermionic and bosonic observables. There would appear to be considerable scope for theoretical studies to flesh out and extend this picture, as well as for testing it in experiment}. 

{\it Note added in proof}: After acceptance of this work, a related preprint appeared~\cite{nowak2021quantum}, which points out that if dephasing at $g=0$ was the only mechanism
of exposing the Gaussian decay of the kink-kink correlation, then it would remain hidden since
there is no dephasing at $g=0$ (the spectrum being flat at that particular point). This does not
contradict our results since in our case, sufficient dephasing can already take place during the
ramp before reaching $g=0$, as is evident from our exact numerical results [Fig.~\ref{fig1} (a)].


\begin{acknowledgments}
The authors thank B. Doyon for a useful discussion,  and W. H. Zurek for very encouraging comments on the manuscript. We especially thank B. Damski for critical reading of the manuscript and many useful suggestions. K.R acknowledges sponsorship, in part, by the Swedish Research Council. K.R and A.D acknowledge the Visitors Program of MPI-PKS for hospitality during a visit during which a part of the project was carried out. A.D acknowledges the partner group program ``Spin liquids: correlations, dynamics and disorder" between IACS and MPI-PKS. This  work  was in  part  supported  by  the  Deutsche  Forschungsgemein-schaft under grants SFB 1143 (project-id 247310070) and the cluster of excellence ct.qmat (EXC 2147, project-id 39085490). 
\end{acknowledgments}

\appendix

\section{Transverse kink correlation}\label{app:A}

The (connected) kink correlation function in the transverse direction is
\begin{align}
 \chi(r) &= \langle (1-\sigma_j^z \sigma_{j+1}^z)(1-\sigma_{j+r}^z \sigma_{j+r+1}^z) \rangle \nonumber \\
 &~~~~~~~~~~~~~~~~~~~~~~~~~~~~ - \langle1-\sigma_j^z \sigma_{j+1}^z\rangle\langle1-\sigma_{j+r}^z \sigma_{j+r+1}^z\rangle \nonumber \\
 &= \langle \sigma_{j}^{z} \sigma_{j+1}^{z} \sigma_{j+r}^z \sigma_{j+r+1}^z \rangle - \langle \sigma_{j}^{z} \sigma_{j+1}^{z} \rangle \langle \sigma_{j+r}^{z} \sigma_{j+r+1}^{z} \rangle \nonumber \\
 &= \langle \mu_{j}^{x} \mu_{j+2}^{x} \mu_{j+r}^x \mu_{j+r+2}^x \rangle - \langle \mu_{j}^{x} \mu_{j+2}^{x} \rangle \langle \mu_{j+r}^{x} \mu_{j+r+2}^{x} \rangle,
 \label{corrtr1}
\end{align}
where in the last line, the duality relations are invoked. 
Introducing fermionic operators $A_j=c_j^\dagger+c_j$ and $B_j=c_j^\dagger-c_j$, making use of the identity ${\rm Exp}[i\pi c_j^\dagger c_j]=A_jB_j$, and denoting $\langle A_jB_{j+r}\rangle \equiv {G}(r)$, we find 
\begin{align}
 \chi(r) &=  \begin{vmatrix}
    G(-1) & G(0) & G(r-1) & G(r) \\ 
    G(-2) & G(-1) & G(r-2) & G(r-1) \\ 
    G(-r-1) & G(-r) & G(-1) & G(0) \\ 
    G(-r-2) & G(-r-1) & G(-2) & G(-1) \\
 \end{vmatrix} \nonumber \\
 &~~~~~~~~~~~~~~~~~~~~~~~~~~~~~~~~~~~~~~~ -
 \begin{vmatrix}
    G(-1) & G(0)  \\ 
    G(-2) & G(-1) \\ 
 \end{vmatrix}^2,
\label{corrtr2}
\end{align}
where we have used Wick contraction of a string of operators of the form $\langle B_jA_{j+1}B_{j+1}A_{j+2} B_{j+r}A_{j+r+1}B_{j+r+1}A_{j+r+2} \rangle$. Note only terms like $A_jB_{j+r}$ survive the contraction (terms like $A_jA_{j+r}$ and $B_jB_{j+r}$ vanish). It then remains to calculate $G(r)$ and using the formulation described in the previous section, we obtain
\begin{align}
 \chi(r) \approx -\frac{\zeta}{4\pi^2\tau_Q} e^{-r^2/2\pi\tau_Q},
\end{align}
where $\zeta=1$ for ramps ending at $g=0$ and $\zeta=1+(-1)^r$ for ramps ending at $g=-\infty$.

\section{Kink correlation in the eigenstates}\label{app:B}

The asymptotic behavior of $G(r)$ [or $G(-r)$] for large $r$ is the following. A straightforward integration yields
\begin{widetext}
\begin{align}
 G(r) = \frac{\pi}{|1+g|}\bigg[ {}_{3}{\cal F}_{2}\big({1/2, 1/2, 1}; {1 - r, 1 + r}; z\big) - g~{}_{3}{\cal F}_{2}\big({1/2, 1/2, 1}; {2 - r, r}; z\big) \bigg]\equiv \pi\frac{{\cal A}-g{\cal B}}{|1+g|},
 \label{hyp1}
\end{align}
\end{widetext}
where $z=4 g/(1 + g)^2$ ($0\leq|z|\leq1$). The quantity ${}_{p}{\cal F}_{q}$ denotes the {\it regularized generalized Hypergeometric function} of the form
\begin{align}
 &{}_{p}{\cal F}_{q}(a_1,\dots, a_p; b_1,\dots, b_q; z) = \sum_{k=0}^\infty T(k) z^k,
 \label{hyp2a}
\end{align}
where
\begin{align}
 &T(k) = \frac{1}{k!}\frac{(a_1)_k\dots (a_p)_k}{(b_1)_k\dots (b_q)_k}\frac{1}{\Gamma(b_1)\dots\Gamma(b_q)},
 \label{hyp2}
\end{align}
$(a_l)_k$ [or $(b_l)_k$] is called the rising factorial or {\it Pochhammer symbol} defined as $(a_l)_k = \Gamma(a_l + k)/\Gamma(a_l)$. Since the hypergeometric series is unimodal, the ratio of two consecutive terms in, e.g., ${\cal A}$ in Eq.~\ref{hyp2a}, which has the form
\begin{align}
 R_{\cal A}\equiv\frac{T(k+1)z^{k+1}}{T(k)z^k} &= \bigg(\frac{z}{k+1}\bigg) \bigg(\frac{\Gamma^2(k+3/2)\Gamma(k+2)}{\Gamma^2(k+1/2)\Gamma(k+1)}\bigg) \nonumber \\
 &~~~~~~~~ \times \bigg(\frac{\Gamma(k+1-r)\Gamma(k+1+r)}{\Gamma(k+2-r)\Gamma(k+2+r)}\bigg) \nonumber \\
                     &= \bigg(\frac{z}{k+1}\bigg) \bigg( \frac{(k+1/2)^2 (k+1)}{(k+1-r)(k+1+r)} \bigg)
 \label{hyp3}
\end{align}
(using the recurrence relation of the Gamma functions $\Gamma(z+1)=z\Gamma(z)$ in the last line), should be equal to $1$ for $k=k_{\rm max}\equiv k_0$. At large values of $r$, we find $k_0\approx r/\sqrt{1-z}$ which also holds for ${\cal B}$. Exploiting its unimodular nature, the series ${\cal A}$ can be approximated by an integral of the form 
\begin{align}
 {\cal A} &\approx \int_{-\infty}^{\infty} {\rm d}k~e^{k{\rm ln}(z)+f(k)} \nonumber \\
 &\approx \int_{-\infty}^{\infty} {\rm d}k~e^{k_0{\rm ln}(z)+f(k_0) + (k-k_0)^2 f''(k_0)/2},
  \label{hyp4}
\end{align}
where we have Taylor expanded the function $f(k)\equiv {\rm ln}[T(k)]$ around $k=k_0$ retaining terms up to quadratic order and using the fact $f'(k_0)=0$. Noting 
\begin{align}
 f(k) \approx {\rm ln}\bigg[ \frac{1}{k!} \frac{\Gamma^3(k)}{\Gamma(k+r)\Gamma(k-r)} \bigg],
\end{align}
and invoking Stirling's approximation of the logarithm of Gamma function and logarithm of factorial, and following the same procedure for ${\cal B}$, we finally obtain
\begin{align}
 G(r) \sim e^{-r/\xi_0},
\end{align}
where
\begin{align}
 \xi^{-1}_0 = |{\rm ln}(4g)-2{\rm ln}(1+g)|+2\sqrt{(1-g)^2/(1+g)^2}.
\end{align}
The correlation length $\xi_0$ expectedly diverges at the critical point $g_c=1$. 

Finally, let us note that $\rho(r)$ in a mixed state, characterized by the occupation probability $n_k=1-\Theta(k-k^\ast)$ for some $k^\ast\in[0,\pi]$ (we call this state a ``box eigenstate'' because of the profile of $n_k$), does not show any Gaussian behavior in the kink correlation. Likewise, when we generate a family of random eigenstates by exciting a random set of $k$ modes with a given probability and measure the kink correlation in those states, we find no trace of Gaussian in any of such states establishing the fact that no eigenstate of ${\cal H}$ can feature the Gaussian behavior, and hence, what we are observing for the ramp is truly a dynamical effect.

\section{Spin correlations}\label{app:C}

Following Ref.~\onlinecite{Sei-Book}, the spin correlation or equivalently the (nonlocal) fermionic correlation function is
\begin{align}
 C_{\rm F}(r) &= \langle \sigma^x_j\sigma^x_{j+r}\rangle = \langle B_j (A_{j+1}B_{j+1}\dots A_{j+r-1}B_{j+r-1}) A_{j+r}\rangle \nonumber \\
 &=\begin{vmatrix}
    G(-1) && G(-2) && \dots && G(-r)   \\
    G(0)  && G(-1) && \dots && G(-r+1) \\
    \dots && \dots && \dots && \dots   \\
    G(r-2) && G(r-3) && \dots && G(-1) \\
   \end{vmatrix},
   \label{fermcorr}
\end{align}
where $G(r)\equiv \langle A_jB_{j+r} \rangle=(1/\pi)\int^\pi_0 {\rm d}k~\big(|u_k|^2-|v_k|^2\big)\cos(kr)$ is the relevant two-point fermionic correlator. The behavior of $C_{\rm F}(r)$ against $r$ for rapid ramps at various (small) values of $\tau_Q$ features oscillations with an exponentially decaying envelope~\cite{Cherng_Levitov}. Furthermore, approximating the quantity $\big(|v_k|^2-|u_k|^2\big)=1-2r_k^2$ by a thermal occupancy of the form $(1+e^{\varepsilon_k/T})^{-1}$ leads to the expressions of $\varepsilon_k$ noted in the main text. From the expression $r_k^2\approx e^{-\pi \tau_Qk^2}$ the effective temperature comes out to be $T=1/(\pi\tau_Q)$ and the critical temperature $T^\ast=1/(\pi\tau_Q^*)$ where $\tau_Q^\ast={\rm ln}(2)/\pi$ as noted in Ref.~[\onlinecite{Cherng_Levitov}]. 

It is to be stressed that the nonlocal fermions behave in a different way than the local fermions (or local bosons like our kinks) at high temperatures because of the presence of the string that gives rise to the Toeplitz determinant noted in Eq.~\ref{fermcorr} above. 
So, even if the quantity $G(r)$ can be approximated by a Gaussian at high temperatures, the determinant would still give rise to a falling exponential implying the spatial correlation of the nonlocal fermions would not display the intuitive Gaussian behavior expected for hot local fermions or bosons that behave like classical (Maxwellian) particles at high temperatures.

\section{Post-ramp correlations}\label{app:D}

Let us assume we have stopped the ramp at $g=g_0$ that yields the wavefunction $[u_k^{(I)}, v_k^{(I)}]^T$ which then evolves to $[u_k(t), v_k(t)]^T$ under the static Hamiltonian $\tilde{\cal H}(g)$ at $g=g_0$. The dynamics is governed by the Schr{\"o}dinger equation
\begin{align}
 i\partial_t
 \begin{pmatrix}
  u_k \\
  v_k
 \end{pmatrix}
 =
 \begin{pmatrix}
 1-g_0\cos k & -ig_0\sin k \\
 ig_0\sin k & g_0\cos k - 1
 \end{pmatrix}
 \begin{pmatrix}
  u_k \\
  v_k
 \end{pmatrix}
 \equiv \tilde{\cal H}_k
  \begin{pmatrix}
  u_k \\
  v_k
 \end{pmatrix},
\label{eom1}
\end{align}
where time dependence in $u_k, v_k$ is implied. If the unitary matrix $U_k$ diagonalizes $\tilde{\cal H}_k$ such that $U_k^\dagger \tilde{\cal H}_k U_k=\Sigma_k$ where $\Sigma_k=
\varepsilon_k
 \begin{pmatrix}
   -1 & 0 \\                                                                                                                                                             
    0 & 1 
  \end{pmatrix}$ 
with $\varepsilon_k=\sqrt{1-2g_0\cos k+g_0^2}$, then the diagonal modes $\tilde{u}_k, \tilde{v}_k$ satisfy
\begin{align}
 i\partial_t
 \begin{pmatrix}
  \tilde{u}_k \\
  \tilde{v}_k
 \end{pmatrix}
 =
 \begin{pmatrix}
 -\varepsilon_k & 0  \\
  0 & \varepsilon_k
 \end{pmatrix}
 \begin{pmatrix}
  \tilde{u}_k \\
  \tilde{v}_k
 \end{pmatrix}.
\label{eom2}
\end{align}
These diagonal modes are connected with the original modes as 
\begin{align}
 \begin{pmatrix}
  \tilde{u}_k \\
  \tilde{v}_k
 \end{pmatrix}
 =
 U_k^\dagger
 \begin{pmatrix}
  {u}_k \\
  {v}_k
 \end{pmatrix}
 =
  \begin{pmatrix}
  -\sin(\theta_k/2) & -i\cos(\theta_k/2) \\
  \cos(\theta_k/2) & -i\sin(\theta_k/2)
 \end{pmatrix}
  \begin{pmatrix}
  {u}_k \\
  {v}_k
 \end{pmatrix},
\label{eom3}
\end{align}
where $\cos\theta_k=(1-g_0\cos k)/\varepsilon_k$. If $\tilde{u}_k, \tilde{v}_k$ are specified by their initial values $\tilde{u}_k^{(\rm I)}, \tilde{v}_k^{(\rm I)}$, then their values at time $t$ (assuming the ramp stopped at $t=0$ when the subsequent evolution starts) is 
\begin{align}
 \tilde{u}_k(t) = e^{i\varepsilon_k t}\tilde{u}_k^{(\rm I)}~~~;~~~ \tilde{v}_k(t) = e^{-i\varepsilon_k t}\tilde{v}_k^{(\rm I)}.
\end{align}
Noting $[\tilde{u}_k^{(\rm I)}, \tilde{v}_k^{(\rm I)}]^T=U_k^{\dagger}[{u}_k^{(\rm I)}, {v}_k^{(\rm I)}]^T$, we obtain the solution $u_k, v_k$ as
\begin{align}
 \begin{pmatrix}
  {u}_k \\
  {v}_k
 \end{pmatrix}
 =
 U_k e^{-i\Sigma_k t} U_k^{\dagger}
 \begin{pmatrix}
  {u}_k^{(\rm I)} \\
  {v}_k^{(\rm I)}
 \end{pmatrix}.
\label{eom5}
\end{align}
Written out explicitly and ignoring all the rapidly oscillating terms as done before, this leads to, at late times, the expressions for $u_k$ and $v_k$ noted in Eq.~\ref{ukvkpostramp}.

\section{Density matrix in the post-ramp GGE}\label{app:E}

Calculations in this appendix are done following Ref.~\onlinecite{vidmar2016generalized}. The formulation applies for a quench from an initial state specified by $g=g_{\rm I}$ to a final state specified by $g=g_{\rm F}$.
In momentum space, the Hamiltonian $\tilde{{\cal H}}$ reads 
\begin{align}
 \tilde{{\cal H}}(g)=\sum_k
 \begin{pmatrix}
  c_k^\dagger & c_{-k}
 \end{pmatrix}
 \begin{pmatrix}
 1-g\cos k & -ig\sin k \\
 ig\sin k & g\cos k-1
 \end{pmatrix}
 \begin{pmatrix}
  c_k \\
  c_{-k}^\dagger
 \end{pmatrix}.
\label{ham1}
\end{align}
The Bogoliubov-de Gennes transformation $c_k = u_k \gamma_k + v_{-k}^* \gamma_{-k}^\dagger$ brings it to a diagonal form
\begin{align}
 \tilde{{\cal H}}(g)=\sum_k
 \begin{pmatrix}
  \gamma_k^\dagger & \gamma_{-k}
 \end{pmatrix}
 \begin{pmatrix}
 \epsilon_k & 0 \\
  0 & -\epsilon_k
 \end{pmatrix}
 \begin{pmatrix}
  \gamma_k \\
  \gamma_{-k}^\dagger
 \end{pmatrix},
\label{ham2}
\end{align}
where $\epsilon_k=\sqrt{1-2g\cos k+g^2}$.
As the Hamiltonian in Eq.~\ref{ham1} is an ensemble of two level systems in $k$-space, the initial state (before the quench) can be written as 
\begin{equation}
 |\Psi_{\rm I}\rangle = |r_{k_1}, r_{-k_1}\rangle\otimes\dots\otimes|r_{k_j}, r_{-k_j}\rangle\otimes\dots.
 \label{initial}
\end{equation}
This state could be an eigenstate of the Hamiltonian at $g=g_{\rm I}$ or one obtained from the ramp stopped at $g=g_{\rm I}$. Time evolution of this state is dictated by the final Hamiltonian $H_{\rm F}$ (with $g=g_{\rm F}$) as
\begin{align}
 |\Psi(t)\rangle = e^{-iH_{\rm F}t}|\Psi_{\rm I}\rangle &= \sum_n e^{-iE_{n}t} |n\rangle\langle n|\Psi_{\rm I}\rangle \nonumber \\
 &=  \sum_n e^{-iE_{n}t} |n\rangle c_n,
 \label{timeevol}
\end{align}
where $c_n$ is the overlap of the initial state with the $n$-th eigenstate of the final Hamiltonian which can also be written as
\begin{align}
 |n\rangle = |p^{[n]}_{k_1}, p^{[n]}_{-k_1}\rangle\otimes\dots\otimes|p^{[n]}_{k_j}, p^{[n]}_{-k_j}\rangle\otimes\dots,
 \label{eigen1}
\end{align}
where $|p^{[n]}_{k}, p^{[n]}_{-k}\rangle$ denotes the occupation of Bogoliubov fermions with $k$ and $-k$ in the $n$-th eigenstate (in a given parity sector) while each $\{k, -k\}$ subspace is spanned by the four vectors $\{|0, 0\rangle, |1, 1\rangle, |1, 0\rangle, |0, 1\rangle\}$. The occupation of the Bogoliubov fermions $\gamma_k^\dagger\gamma_k$ are conserved quantities (denoted further, $J_k$).

Observables in integrable systems are expected to relax according to a GGE which is defined by the density matrix 
\begin{align}
 \varrho_{\rm GGE} = Z^{-1} e^{-\sum_k \lambda_k J_k},
 \label{dm1}
\end{align}
where $Z$ is the partition function given by $Z={\rm Tr}[e^{-\sum_k \lambda_k J_k}]$ and $J_k$ denotes the conserved quantities. The Lagrange multipliers are fixed by the condition $\langle \Psi_{\rm I}|J_k |\Psi_{\rm I}\rangle \equiv \langle J_k \rangle_{\rm I} = {\rm Tr}[\varrho_{\rm GGE} J_k]$ which leads to 
\begin{align}
 \lambda_k = {\rm ln}\bigg[\frac{1-\langle J_k \rangle_{\rm I}}{\langle J_k \rangle_{\rm I}}\bigg].
 \label{lm}
\end{align}

One can write $\varrho_{\rm GGE}$ as a sum over the contribution from all the eigenstates of the Hamiltonian
\begin{align}
 \varrho_{\rm GGE} = \sum_n \varrho^{[n]}_{\rm GGE} |n\rangle \langle n|,
\end{align}
where 
\begin{align}
 \varrho^{[n]}_{\rm GGE} = \prod_k^{p^{[n]}_{k}=0} \big( 1- \langle J_k \rangle_{\rm I} \big) \prod_k^{p^{[n]}_{k}= 1} \langle J_k \rangle_{\rm I}, 
 \label{dm2}
\end{align}
using $\langle n|e^{-\lambda_k J_k}|n\rangle=\delta_{p^{[n]}_{k},0}+e^{-\lambda_k}\delta_{p^{[n]}_{k},1}$ (both $k$ and $-k$ are included in the above product). To complete the discussion, what remains is to compute $\langle J_k \rangle_{\rm I}$.

Calculating $\langle J_k \rangle_{\rm I}$ involves the overlap $c_n$ in Eq.~\ref{timeevol} which is the product of the overlaps (for a chain of length $N$, it is a product over $N/2$ terms in each of the parity sectors) in each subspace $\{k, -k\}$ as $c_n=\prod_k c_k^{[n]}$. Denoting the initial state as a tensor product $|\Psi_{\rm I}\rangle=\otimes_k \big[ u_k^{(\rm I)}~~v_k^{(\rm I)}\big]^T$ and that corresponding to the final state $|\Psi_{\rm F}\rangle=\otimes_k \big[u_k^{(\rm F)}~~v_k^{(\rm F)}\big]^T$, we have
\begin{align}
 c_k^{[n]} = \begin{cases}
  \pm \sqrt{\alpha_k} ~~~~~~~~\equiv c_k^{(1)}, & \text{if } p^{[n]}_{k}=r_k~~\text{and}~~p^{[n]}_{-k}=r_{-k}, \\
  0 ~~~~~~~~~~~~~~~~~\equiv c_k^{(2)}, & \text{if } p^{[n]}_{k}=1~~~~\text{and}~~p^{[n]}_{-k}=0, \\
  0 ~~~~~~~~~~~~~~~~~\equiv c_k^{(3)}, & \text{if } p^{[n]}_{k}=0~~~~\text{and}~~p^{[n]}_{-k}=1, \\
  \pm i\sqrt{1-\alpha_k} \equiv c_k^{(4)}, & \text{if } p^{[n]}_{k}\neq r_k~~\text{and}~~p^{[n]}_{-k}\neq r_{-k},
\end{cases}
\label{overlap1}
\end{align} 
for $r_k=r_{-k}=0$ and $r_k=r_{-k}=1$, and
\begin{align}
 c_k^{[n]} = \begin{cases}
  0 \equiv c_k^{(1)}, & \text{if } p^{[n]}_{k}=0~~~~\text{and}~~p^{[n]}_{-k}=0, \\
  1 \equiv c_k^{(2)}, & \text{if } p^{[n]}_{k}=r_k~~\text{and}~~p^{[n]}_{-k}=r_{-k}, \\
  0 \equiv c_k^{(3)}, & \text{if } p^{[n]}_{k}\neq r_k~~\text{and}~~p^{[n]}_{-k}\neq r_{-k}, \\
  0 \equiv c_k^{(4)}, & \text{if } p^{[n]}_{k}=1~~~~\text{and}~~p^{[n]}_{-k}=1,
\end{cases}
\label{overlap2}
\end{align} 
for $r_k=0$ and $r_{-k}=1$, or, $r_k=1$ and $r_{-k}=0$. Note $\sum_{\xi}|c_k^{(\xi)}|^2=1$ is implied by the normalization of the initial state $\langle\Psi_{\rm I}|\Psi_{\rm I}\rangle=1$.

Using the notation in Eq.~\ref{overlap1} and Eq.~\ref{overlap2} we find $\langle J_k \rangle_{\rm I}=|c_k^{(\xi)}|^2$ with $p^{[n]}_{k}=1$ while $\xi$ is determined by $\{r_k,r_{-k}\}$ and so
\begin{align}
 \langle J_k \rangle_{\rm I} = \begin{cases}
  1-\alpha_k, & \text{if } r_k=0~~\text{and}~~r_{-k}=0, \\
  \alpha_k, & \text{if } r_k=1~~\text{and}~~r_{-k}=1, \\
  0, & \text{if } r_k=1~~\text{and}~~r_{-k}=0, \\
  0, & \text{if } r_k=0~~\text{and}~~r_{-k}=1,
\end{cases}
\label{overlap3}
\end{align}
where 
\begin{align}
 \alpha_k = \big|u_k^{(\rm I)*}u_k^{(\rm F)}+v_k^{(\rm I)*}u_k^{(\rm F)}\big|^2.
 \label{alphak}
\end{align}

As the initial state prepared at the end of the ramp obeys $\gamma_{k}\gamma_{-k}|\Psi_{\rm I}\rangle =0$, it corresponds to $r_k = r_{-k} = 0$ in Eq.~\ref{initial} implying $\langle J_k \rangle_{\rm I} =1-\alpha_k$ and
\begin{align}
 \varrho^{[n]}_{\rm GGE} = \prod_k^{p^{[n]}_{k}=0} \alpha_k^2 \prod_k^{p^{[n]}_{k}\neq p^{[n]}_{-k}} \alpha_k(1-\alpha_k) \prod_k^{p^{[n]}_{k}=1} (1-\alpha_k)^2.
 \label{dm3}
\end{align}

\section{Kink correlation in a GGE}\label{app:F}

As noted in the main text, the (longitudinal) kink correlator is 
\begin{align}
\hat{\rho}(r) = \mu_j^z\mu_{j+r}^z - \langle\mu_j^z\rangle^2.
\end{align}
Introducing fermionic operators $A_j=c_j^\dagger+c_j$ and $B_j=c_j^\dagger-c_j$, we can write $\mu_j^z\equiv 1-2c_j^\dagger c_j=A_jB_j$. Further denoting $A_jB_{j+r}\equiv \hat{G}(r)$, we find
\begin{align}
 \rho_{\rm GGE}(r)\equiv\langle \hat{\rho}(r) \rangle_{\rm GGE} = -\sum_n \varrho^{[n]}_{\rm GGE} \langle n|\hat{G}(r)|n\rangle \langle n|\hat{G}(-r)|n\rangle,
 \label{corrgge}
\end{align}
where $\langle n|\hat{G}(r)|n\rangle$, written in terms of the occupation of the Bogoliubov fermions, is
\begin{align}
 \langle n|\hat{G}(r)|n\rangle = \frac{2}{N} \sum_{k>0} (p^{[n]}_{k}+p^{[n]}_{-k}-1) \mathcal{G},
 \label{corrGr}
\end{align}
with
\begin{align}
 \mathcal{G} = \frac{g\sin k \sin(kr)}{\sqrt{1-2g\cos k+g^2}} - \frac{(1-g\cos k) \cos(kr)}{\sqrt{1-2g\cos k+g^2}}.
 \label{Gkr}
\end{align}

Expectation value of an operator $\hat{\mathcal{A}}$ in the GGE with weights $\{\varrho_{\rm GGE}^{[n]}\}$ is given by
\begin{align}
 \langle \hat{\mathcal{A}} \rangle_{\rm GGE} = \sum_n \varrho^{[n]}_{\rm GGE} \langle n|\hat{\mathcal{A}}|n\rangle.
\label{expvl1} 
\end{align}
Provided $\langle n|\hat{\mathcal{A}}|n\rangle$ is a sum of single particle contribution as $\langle n|\hat{\mathcal{A}}|n\rangle=\sum_k \langle n|\hat{\mathcal{A}}_k|n\rangle$ and the weight $\varrho_{\rm GGE}^{[n]}$ factorizes as $\varrho_{\rm GGE}^{[n]}=\prod_k \varrho_{k,{\rm GGE}}^{[n]}$, Eq.~\ref{expvl1} can be rewritten as
\begin{align}
  \langle \hat{\mathcal{A}} \rangle_{\rm GGE} = \sum_{k>0} \bigg(\sum_{\xi} \varrho^{(\xi)}_{k,{\rm GGE}} \hat{\mathcal{A}}_k^{(\xi)}\bigg),
\label{expvl2}
\end{align}
where we have introduced the notation
\begin{align}
 \varrho^{[n]}_{k,{\rm GGE}} = \begin{cases}
  \varrho^{(1)}_{k,{\rm GGE}}, & \text{if } p^{[n]}_{k}=0~~\text{and}~~p^{[n]}_{-k}=0, \\
  \varrho^{(2)}_{k,{\rm GGE}}, & \text{if } p^{[n]}_{k}=1~~\text{and}~~p^{[n]}_{-k}=0, \\
  \varrho^{(3)}_{k,{\rm GGE}}, & \text{if } p^{[n]}_{k}=0~~\text{and}~~p^{[n]}_{-k}=1, \\
  \varrho^{(4)}_{k,{\rm GGE}}, & \text{if } p^{[n]}_{k}=1~~\text{and}~~p^{[n]}_{-k}=1,
\end{cases}
\label{rho}
\end{align}
and
\begin{align}
 \langle n|\hat{\mathcal{A}}_k|n\rangle = \begin{cases}
  \hat{\mathcal{A}}_k^{(1)}, & \text{if } p^{[n]}_{k}=0~~\text{and}~~p^{[n]}_{-k}=0, \\
  \hat{\mathcal{A}}_k^{(2)}, & \text{if } p^{[n]}_{k}=1~~\text{and}~~p^{[n]}_{-k}=0, \\
  \hat{\mathcal{A}}_k^{(3)}, & \text{if } p^{[n]}_{k}=0~~\text{and}~~p^{[n]}_{-k}=1, \\
  \hat{\mathcal{A}}_k^{(4)}, & \text{if } p^{[n]}_{k}=1~~\text{and}~~p^{[n]}_{-k}=1.
\end{cases}
\label{op}
\end{align}
Similarly, it is straightforward to show the identity
\begin{align}
  &\sum_n \varrho^{[n]}_{k,{\rm GGE}}\langle n|\hat{\mathcal{A}}_1|n\rangle \langle n|\hat{\mathcal{A}}_2|n\rangle = 
  \langle \hat{\mathcal{A}}_1 \rangle_{\rm GGE}\langle \hat{\mathcal{A}}_2 \rangle_{\rm GGE}
  \nonumber \\ 
  &-\sum_{k>0}\bigg( \sum_\xi\varrho^{(\xi)}_{k,{\rm GGE}} \hat{\mathcal{A}}_{1,k}^{(\xi)} \bigg)\bigg( \sum_{\xi'}\varrho^{(\xi')}_{k,{\rm GGE}} \hat{\mathcal{A}}_{2,k}^{(\xi')} \bigg) \nonumber \\ &+\sum_{k>0} \bigg(  \sum_\xi \varrho^{(\xi)}_{k,{\rm GGE}} \hat{\mathcal{A}}_{1,k}^{(\xi)}\hat{\mathcal{A}}_{2,k}^{(\xi)}\bigg),
\label{expvl3} 
\end{align}
where $\langle \hat{\mathcal{A}}_{1/2} \rangle_{\rm GGE}$ is defined according to Eq.~\ref{expvl2} and $\hat{\mathcal{A}}_{1/2,k}^{(\xi)}$ is defined according to Eq.~\ref{op}. Using Eq.~\ref{dm3} to Eq.~\ref{expvl3}, we obtain the expression for ${\rho}_{\rm GGE}(r)$ noted in Eq.~\ref{rhoGGE}. 

\section{Pearson correlation coefficient}\label{app:G}

For a paired dataset $\mathcal{D}:=\{(x_i, y_i)\}$, the correlation coefficient ${\mathcal R}$ ($\in[-1,1]$) of the best fit is given by
\begin{equation}
 \mathcal{R}=(\langle{XY}\rangle-\langle{X}\rangle\langle{Y}\rangle)/(\sigma_X~\sigma_Y),
\end{equation}
where $\langle X(Y) \rangle$ denotes the mean of the sample set $X=\{x_i\}$($Y=\{y_i\}$) and $\sigma$, the standard deviation. If $|{\cal R}|=1$, the data is said to have a perfect linear correlation (all points are on the best fit) while $|{\cal R}|=0$ implies no correlation whatsoever.
Accordingly, if the data is perfectly Gaussian correlated, $|\mathcal{R}|$ is $1$ for $X=\{x_i^2\}$ and $Y=\{{\rm ln}|y_i|\}$ and deviation from $|\mathcal{R}|=1$ suggests departure from Gaussian. Following this, we compute $|\mathcal{R}|$ for our numerical data $\mathcal{D}:=\{(r_i^2, {\rm ln}|\rho_i|)\}$ and study its behavior as a function of $\hat{g}$ for different values of $\tau_Q$. 

\bibliographystyle{apsrev4-1}
\bibliography{references}

\end{document}